\documentclass[aps,prl,twocolumn,showpacs,letterpaper, superscriptaddress]{revtex4}
\usepackage{amsmath}
\usepackage{graphicx}
\usepackage{graphics}
\begin{document}

\title{Resolving magnetic frustration in a pyrochlore lattice}
\author{Jiyang Wang}
\affiliation{The James Franck Institute and Department of Physics, The University of Chicago, Chicago, IL 60637, USA}
\author{Yejun Feng}
\email [Corresponding author.  Email: ]  {yejun@aps.anl.gov}
\affiliation{The James Franck Institute and Department of Physics, The University of Chicago, Chicago, IL 60637, USA}
\affiliation{The Advanced Photon Source, Argonne National Laboratory, Argonne, IL 60439, USA}
\author{R. Jaramillo} 
\affiliation{School of Engineering and Applied Sciences, Harvard University, Cambridge, MA 01238, USA}
\author{Jasper van Wezel}
\affiliation{The Materials Science Division, Argonne National Laboratory, Argonne, IL 60439, USA}
\author{P. C. Canfield}
\affiliation{Ames Laboratory and Department of Physics and Astronomy, Iowa State University, Ames, Iowa 50011, USA}
\author{T. F. Rosenbaum} 
\affiliation{The James Franck Institute and Department of Physics, The University of Chicago, Chicago, IL 60637, USA}

\begin{abstract}
CeFe$_2$ is a geometrically frustrated ferromagnet that lies close to an instability at which a subtle change in the lattice symmetry couples to a transition to antiferromagnetism. We use x-ray diffraction, diamond-anvil-cell techniques, and numerical simulation to identify the ground states and to quantitatively illustrate effects of competing magnetic energy scales and geometrical frustration on the magnetic phase diagram. Comparison of phase transitions under both chemical substitution and applied pressure suggests a general solution to the physics of pyrochlore rare earth inter-metallic magnets. 
\end{abstract}



\maketitle

In a geometrically frustrated magnet not all pairs of magnetic ions can simultaneously assume their lowest energy configuration, and the system often has to choose among a large number of competing configurations. From a conceptual standpoint, frustrated systems are interesting because ostensibly minor phenomena can play an outsized role in dictating the overall magnetic structure. CeFe$_2$ is a frustrated rare earth inter-metallic magnet with Laves crystal symmetry. Laves phase materials belong to a broader class of pyrochlore lattice systems, which attract much interest as canonical examples of geometrically frustrated magnetism in three dimensions \cite{1}. Many Laves magnets of both ferro- and antiferromagnetic symmetry are known \cite{4, 5, 6, 7, 8, 9} and, as we demonstrate here, the ease with which the magnetic symmetry can be switched is a direct result of geometrical frustration. In this work we quantitatively illustrate the effects of geometrical frustration on the generic magnetic phase diagram of CeFe$_2$.  

\begin{figure}
\begin{center}
\includegraphics[width=3.375in]{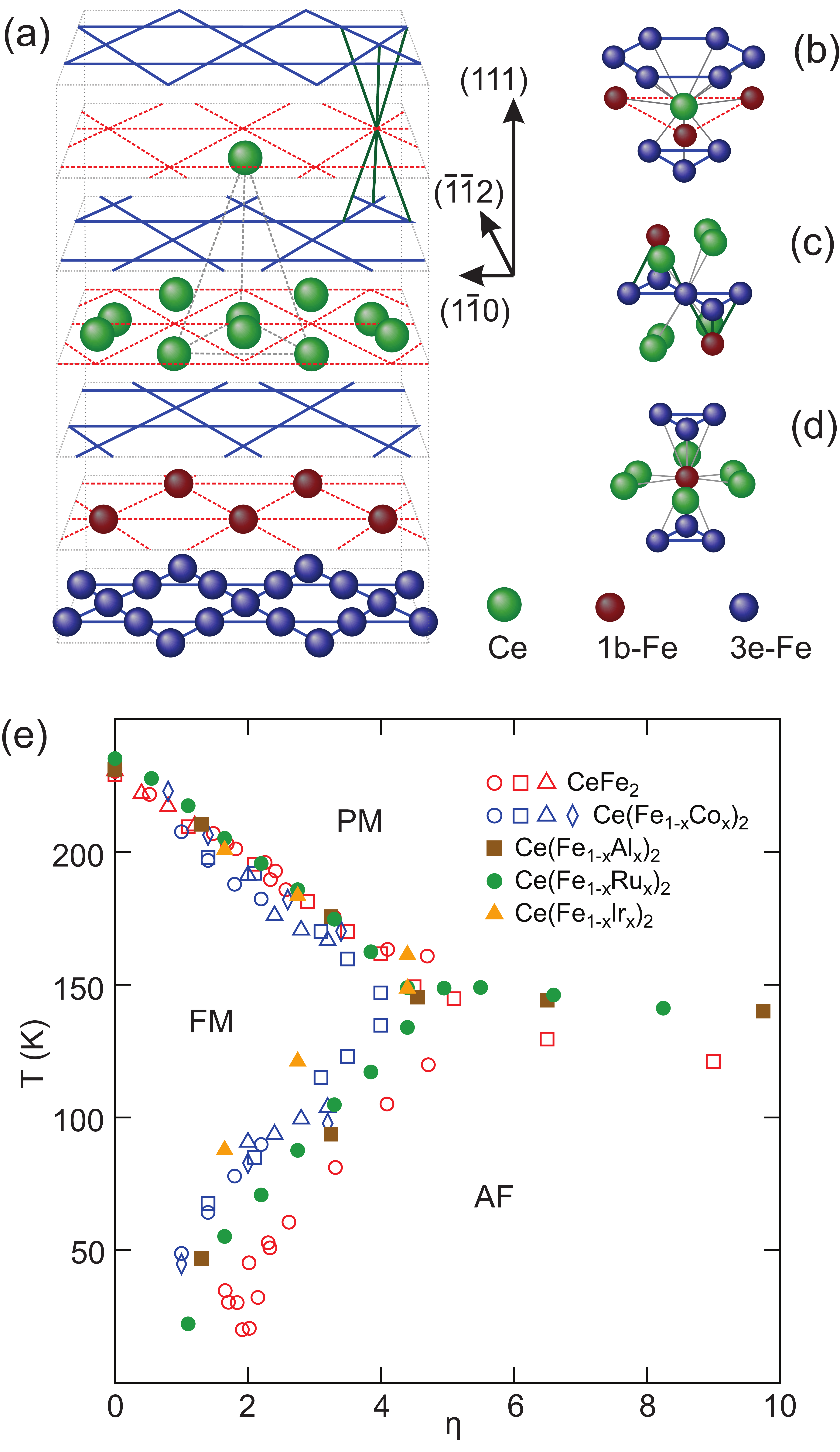}
\caption{(color online). Crystal structure and magnetic phase diagram of CeFe$_2$. (a) Layered view of CeFe$_2$ along ${\langle}111{\rangle}$.  The structure consists of stacked sheets of 3e-Fe sites in a kagome lattice, alternating with corrugated sheets of Ce sites and 1b-Fe sites in a triangular lattice. For clarity, the ${\langle}111{\rangle}$ axis is elongated and not all atoms are shown. (b-d) The twelve nearest neighbors for each site. (e) Generic magnetic phase diagram for CeFe$_2$ showing the paramagnetic (PM), ferromagnetic (FM), and antiferromagnetic (AF) phases. The tuning parameter $\eta$ is a linear combination of pressure P (in GPa) and doping x (in $\%$) according to $\eta = P + Ax$, where $A$ is a dopant-specific scaling factor: $A$ = 20, 65, 55, and 55 GPa/$\%$ for Co, Al, Ru, and Ir, respectively. All four dopants would individually preserve the Laves phase when fully replace Fe in CeFe$_2$ \cite{11}. For pure CeFe$_2$ under pressure and Co-doped CeFe$_2$, the lattice constant shrinks with increasing $\eta$ \cite{12, 13, 14, 29}. In contrast, the lattice expands for doping with Al, Ru and Ir \cite{15,16,17}. }
\label{fig1}
\end{center}
\end{figure}

The cubic Laves structure features transition metal ions on a pyrochlore lattice of corner-sharing tetrahedra woven through by corrugated planes of rare earth ions, as illustrated in Fig.~\ref{fig1}. Stoichiometric CeFe$_2$ \cite{10, 11, 28} is ferromagnetic but undergoes a phase transition to an antiferromagnet under light chemical doping by for example Al, Co, Ru or Ir \cite{12, 13, 14, 15, 16,17}. Previous studies on stoichiometric CeFe$_2$ have found evidence for a high pressure phase, which by comparison with the doped systems was posited to be antiferromagnetic \cite{12}. Further support for this identification comes from the observation of antiferromagnetic fluctuations within the ferromagnetic phase \cite{7, 8, 18}. We show in Fig.~\ref{fig1} that the existing data for chemical doping and applied pressure can be collapsed onto a generic phase diagram, wherein both the ferromagnetic and the antiferromagnetic transition temperatures are plotted as a function of an effective tuning parameter $\eta$. Interestingly, $\eta$ does not vary monotonically with the lattice constant, and the antiferromagnetic phase can be found for materials with lattice constants both larger and smaller than the parent compound CeFe$_2$. This apparent insensitivity to the lattice constant is at variance with the conventional notion that chemical substitution influences magnetism by exerting a positive or a negative ``chemical pressure", and suggests that the phase diagram is instead controlled by symmetry and the geometry of magnetic frustration. 

\begin{figure}
\begin{center}
\includegraphics[width=3.375in]{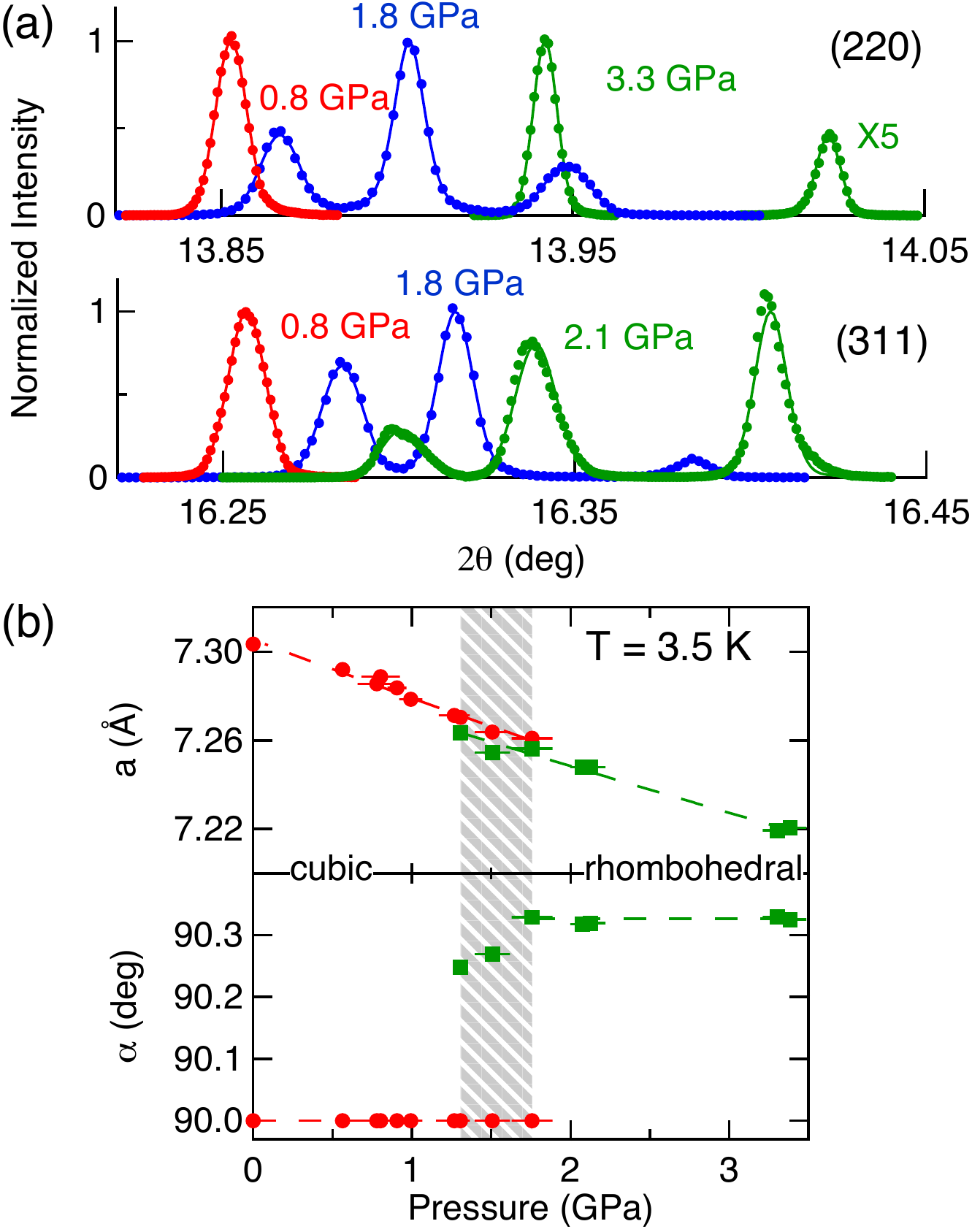}
\caption{(color online).  Evolution of the CeFe$_2$ crystal lattice with pressure at 3.5 K. (a) Longitudinal ($\theta-2\theta$) scans of the (220) and (311) Bragg peaks at three pressures: in the low-pressure cubic phase, in the coexistence regime, and in the high pressure rhombohedral phase. The peak splitting at high pressure is evidence of the rhombohedral distortion.  At 1.8 GPa, the splitting of the (220) peak indicates phase coexistence, while the cubic (311) peak is indistinguishable from the rhombohedral (31$\overline{1}$) peak within our measurement resolution. (b) Dependence of the lattice constant $a$ and the cell-axis angle $\alpha$ on pressure.  The shaded area marks the phase coexistence regime. }
\label{fig2}
\end{center}
\end{figure}

With sensitivity to both lattice and magnetic symmetry, and compatibility with high pressure diamond anvil cells \cite{19, 21}, synchrotron x-ray diffraction is the ideal technique to study the phase diagram of a geometrically frustrated antiferromagnet (Supplemental Material). We plot in Fig.~\ref{fig2} the response of the crystal lattice to applied pressure $P$ at low temperature. Near $P = 1.5$ GPa the lattice undergoes a transition in which it is compressed along one of the cubic body diagonals, resulting in a splitting of the cubic crystal into four types of rhombohedral domains. Using high resolution x-ray diffraction it is possible to index diffraction peaks by their rhombohedral domain type. The compressibility $B_0$ is determined by fitting the lattice constant $a$($P$) to a one parameter Birch equation \cite{21}, with $B_0 = 90 \pm 4$ GPa and $105 \pm 5$ GPa in the low and high pressure phases respectively. In the high pressure phase the unit cell angle $\alpha$ deviates from $90^\circ$ by $0.327 \pm 0.002 ^{\circ}$. The data also show clear evidence for a regime of phase coexistence between 1.3 and 1.8 GPa. The presence of a structural phase transition and a regime of phase coexistence are consistent with magnetic susceptibility studies of CeFe$_2$ under pressure, which found the ferromagnetic phase boundary to be difficult to pin down at low temperature \cite{12}. Note also that a similar rhombohedral distortion is known to accompany the ferromagnetic-to-antiferromagnetic phase transition in Ru-, Al-, and Co-doped CeFe$_2$, with $\alpha$ in the range $90.2^\circ - 90.31^\circ$ \cite{8, 22}, further supporting the notion that the magnetic phase diagram is linked to symmetry considerations rather than controlled by effective chemical pressure or the effects of disorder.

\begin{figure}
\begin{center}
\includegraphics[width=3.375in]{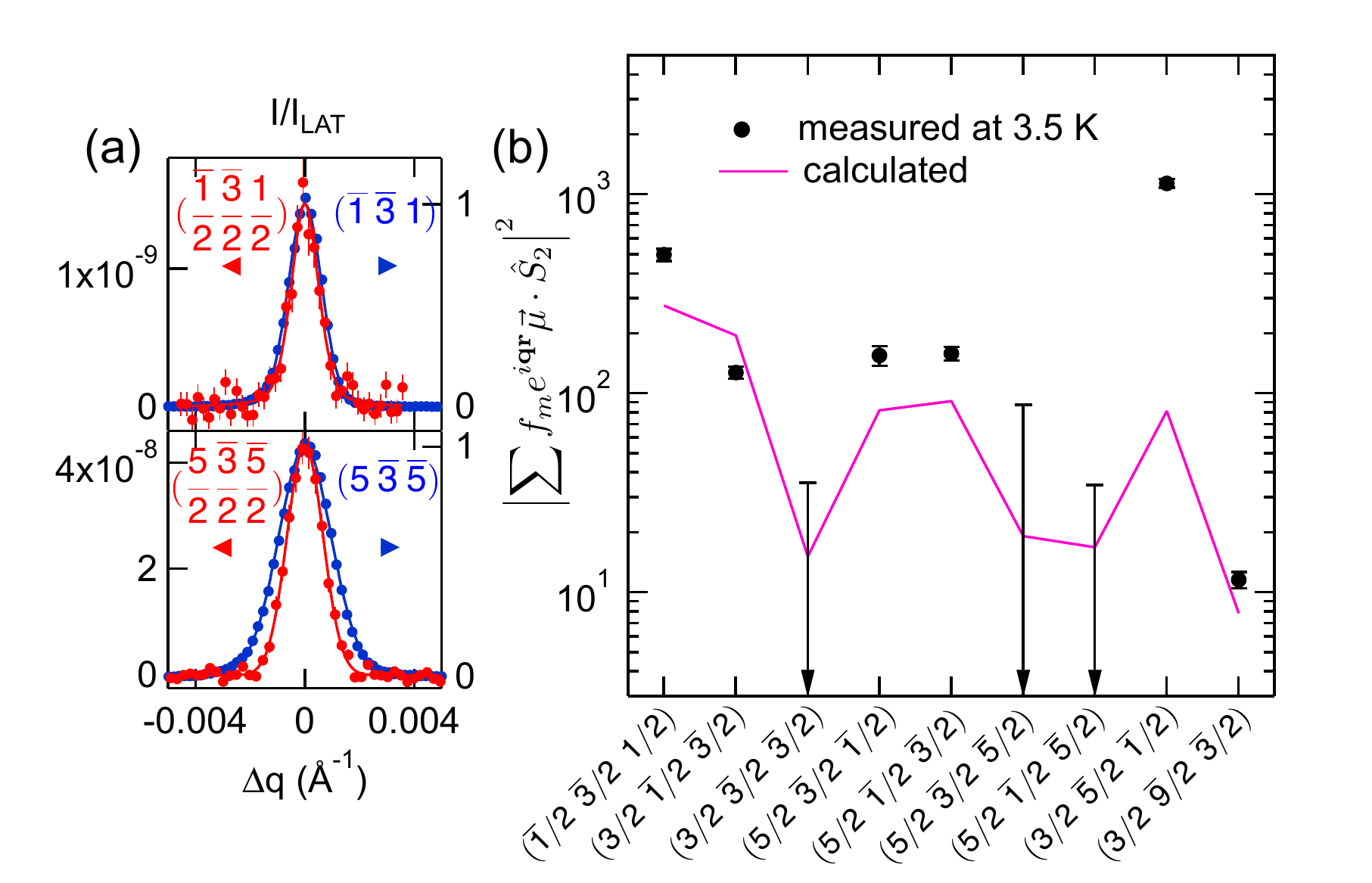}
\caption{(color online). Direct observation of antiferromagnetism at 3.3 GPa and 3.5 K. (a) Longitudinal scans of magnetic peaks (red) normalized to the lattice peaks (blue) respectively. (b) Quantity $|\Sigma f_{m}e^{i\mathbf{qr}}\vec{\mu}\cdot\hat{S}_2|^2$ calculated using Eq.~\eqref{eq1} for six measured diffraction peaks (black circles). Also shown are the sensitivity limits of null measurements at three positions (black error bars with downward arrows). Calculated values (purple line) are given for the model described in the text. }
\label{fig3}
\end{center}
\end{figure}

Using non-resonant magnetic x-ray diffraction we searched for and found the high pressure antiferromagnetic phase in the form of (1/2, 1/2, 1/2)-type Bragg diffraction peaks associated with antiferromagnetic period doubling \cite{20}. The longitudinal magnetic peak widths are limited by the instrument resolution (Fig.~\ref{fig3}a), meaning that the coherence length of the antiferromagnetic domains is at least 1500 $\mathrm{\AA}$. This direct observation of antiferromagnetic order in compressed, stoichiometric CeFe$_2$, not just in its doped analogues, is strong justification for the generic phase diagram drawn in Fig.~\ref{fig1}.

We show in Fig.~\ref{fig3}b antiferromagnetic reflections at nine positions in reciprocal space, all corresponding to a single rhombohedral domain at $P = 3.3$ GPa and $T = 3.5$ K. The nine positions yielded six measurable peaks and three null results, all of which can be used to constrain the magnetic structure. The orbital contributions to the magnetism in CeFe$_2$ are negligible compared to the spin contribution for both Ce and Fe \cite{10}. The non-resonant magnetic diffraction cross section is thus dominated by the projection of the spin onto the direction ($\hat{S}_2$) perpendicular to the scattering plane. The measured quantity is the ratio of diffraction intensity from the antiferromagnetic and the lattice Bragg peaks, expressed as
\begin{align}
\frac{I_{AF}(\frac{h}{2}, \frac{k}{2}, \frac{l}{2})}{I_{LAT}(h, k, l)}=\left(\frac{\hbar\omega}{m_{e}c^2}  \frac{\sum f_{m}e^{i\mathbf{qr}}\vec{\mu}\cdot\hat{S}_2}{8\sum f_e e^{i\mathbf{qr}}} \right)^2 \sin^2(2\theta).
\label{eq1}
\end{align}
Here (h, k, l) are Miller indices of the unit cell before antiferromagnetic period doubling. The summations in the numerator and denominator run over all scattering sites in the magnetic and lattice unit cells, respectively, and the factor of eight accounts for the difference in size between the magnetic and lattice unit cells. The x-ray energy is $\hbar\omega$, while $2\theta$ is the diffraction angle of (h/2, k/2, l/2), and $f_m$ and $f_e$ are the magnetic and atomic form factors \cite{10}. 

\begin{figure}
\begin{center}
\includegraphics[width=3.375in]{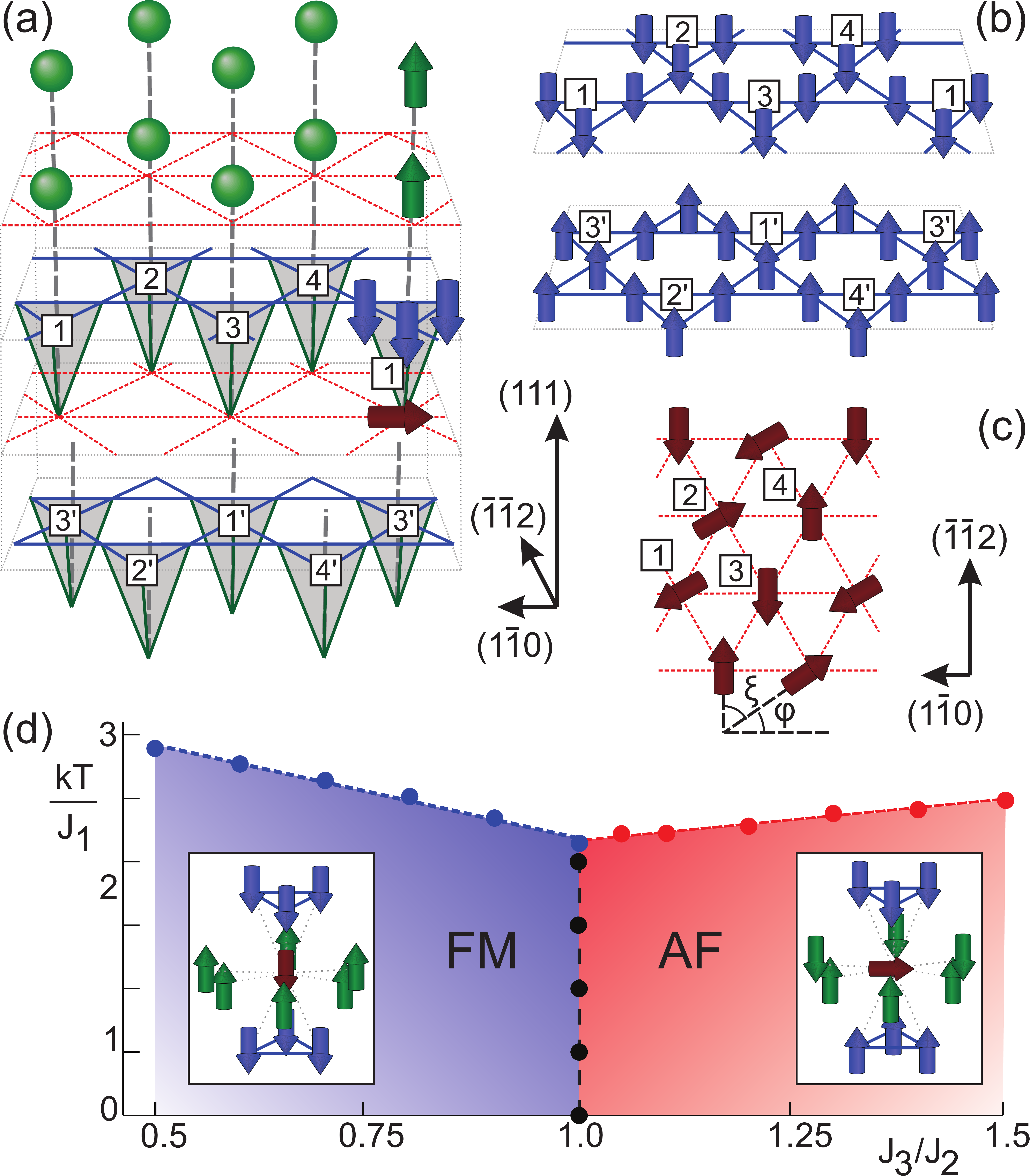}
\caption{(color online). Antiferromagnetic structure of CeFe$_2$. (a) Schematic showing the bipartite sub-lattices (1-4) and (1'-4') for a face centered antiferromagnet \cite{23}. Each sub-lattice point is associated with a complete Laves basis, which is explicitly drawn on sub-lattice point 1.
The spin orientations on site 1' are the inverse of those on site 1, and likewise for the other pairs. (b) The 3e-Fe spins on each kagome sheet are ferromagnetically aligned. (c) The 1b-Fe spins form a plane of frustrated triangular plaquettes. The degeneracy associated with the choice of antiferromagnetic pairs leads to a magnetic domain structure; only one domain is drawn here as an example. (d) Monte Carlo simulation of the generic magnetic phase diagram, which is controlled by competition between the antiferromagnetic Ce-Ce exchange interaction ($J_3$), and the ferromagnetic Fe-Fe interaction ($J_2$). The spin structure for CeFe$_2$ in both the ferromagnetic and antiferromagnetic phases is drawn for the 1b-Fe site and its twelve nearest neighbors.}
\label{fig4}
\end{center}
\end{figure}

In order to constrain the magnetic structure we resort to a general treatment of antiferromagnetic order on a face centered lattice \cite{23}. The ordering doubles the unit cell along all three axes in real space. For a face centered lattice the basis therefore increases from one to eight points, forming a bipartite lattice consisting of basis points 1-4 and 1'-4', with the condition that the primed and unprimed points be magnetically distinguishable. In Fig.~\ref{fig4} we choose the assignment of the basis points that is most natural to a structure layered along $\langle 111 \rangle$. Bear in mind that the Laves basis consists of six atomic sites (four Fe and two Ce) at each point of the bipartite lattice.

The antiferromagnetic structure proposed for Ce(Fe$_{1-x}$Co$_x$)$_2$ on the basis of neutron and resonant x-ray diffraction measurements \cite{8, 9}, consists of ferromagnetic kagome sheets of 3e-Fe spins (each carrying 1.61 $\mu$B) polarized along $\langle111\rangle$, with the spins inverted on adjacent sheets. The Ce spins (0.13 $\mu$B) are also parallel to $\langle111\rangle$ and are ordered antiferromagnetically within each (111) plane. The 1b-Fe spins (1.12 $\mu$B) are polarized within the (111) plane. This can be understood as the result of frustration, arising from the ferromagnetic coupling of the 1b-Fe spins to the two oppositely polarized neighboring kagome sheets. Considering only the coupling of the 1b-Fe spins with their (Ce and 3e-Fe) nearest neighbors, all orientations of the spins within the plane are degenerate. In the model of Ref. \cite{9} the sheets of 1b-Fe spins are ferromagnetically polarized in the plane, so that the lattice sites 1-4 have identical spin orientations, which are opposite to the spins on sites 1'-4'. However, it is straightforward to show that any such model gives a magnetic structure factor of zero for all six of the observed reflections plotted in Fig.~\ref{fig3}. 

The model proposed for Ce(Fe$_{1-x}$Co$_x$)$_2$ is therefore not directly transferrable to antiferromagnetic CeFe$_2$ under pressure. However, this model can conform to our data with one revision. The fact that the Fe spins are not all co-linear is established \cite{9}, and the assumption that the 1b-Fe spin is misaligned with the 3e-Fe spins is consistent with the rhombohedral distortion. However, the interaction between 1b-Fe spins and their orientation within the (111) plane have not been constrained. If this interaction is antiferromagnetic rather than ferromagnetic, then the triangular plaquettes of 1b-Fe sites will be magnetically frustrated, the bipartite lattice sites 1-4 will not have identical spin orientations, and the observed reflections acquire non-zero structure factors. 

As explained in detail in the Supplemental Material, this leads to a model (Fig.~\ref{fig4}) that is consistent with our data. The planes of 1b-Fe sites consist of antiferromagnetically aligned pairs and two spin polarization axes separated by $\xi = 60^\circ$. There is degeneracy associated with the choice of pairs, giving rise to a magnetic domain structure. Each domain contributes to some but not all of the reflections shown in Fig.~\ref{fig3}b and for simplicity we assume that the three domain types are equally populated. We further assume that the angle $\varphi$ in Fig.~\ref{fig4}c differs by $120^\circ$ between domains. The diffraction intensities calculated from this model are in decent quantitative agreement with our data (Fig.~\ref{fig3}b), including the three null results, and allow us to determine the 1b-Fe moment to be $1.80 \pm 0.04 \mu$B. The agreement between model and measurements can be further improved by assuming unequal magnetic domain populations and/or by optimizing the angles $\varphi$ and $\xi$, but we feel such an analysis is unjustified within our limited data set. 

We can now understand the phase diagram by considering the magnetic interactions. In both the high and low pressure phases, the Ce spins are anti-aligned with the majority of the Fe spins \cite{8, 9, 10, 25}. This suggests that an antiferromagnetic coupling (with exchange energy $J_1$) between the Ce and Fe spins is the dominant magnetic interaction. This coupling may be expected to arise from the double exchange interaction between the localized Fe moments and the itinerant Ce electrons \cite{26}, and is known to exist in Laves systems \cite{4, 27}. The bonds between Fe nearest neighbors are ferromagnetic (exchange energy $J_2$). Finally, the Ce spins are antiferromagnetically aligned in the high pressure phase. The proximity of the magnetic phase transition in CeFe$_2$ to ambient pressure, and the insensitivity to the interatomic distances (Fig.~\ref{fig1}), are strong indications that the magnetic interactions will not be \textit{qualitatively} different on opposite sides of the transition. We therefore propose that the Ce-Ce interaction is always antiferromagnetic (exchange energy $J_3$), and that the shifting balance between the antiferromagnetism of the Ce spins and the ferromagnetism of the Fe spins drives the transition. 

To test this proposal we construct a semiclassical model, taking into account only the nearest neighbor interactions $J_1$, $J_2$ and $J_3$. We numerically determine the phase diagram both within a mean field approximation and using a classical Monte Carlo simulation (see Supplementary Material), assuming that the Ce and 3e-Fe spins will be aligned along $\langle111\rangle$. The resulting phase diagram is shown in Fig.~\ref{fig4}d. Even at this level of approximation, the theoretical model clearly reproduces the qualitative features of the experimental phase diagram.

The generic magnetic phase diagram in Fig.~\ref{fig1} is now explained. The low temperature, ambient pressure phase of CeFe$_2$ is ferromagnetic, but it sits close to an antiferromagnetic instability and $J_{3}/J_{2}$ is close to the critical value. Since Al, Co, Ru and Ir all replace Fe upon being doped into CeFe$_2$, they affect the Fe-Fe bonds more strongly than the Ce-Ce bonds. The dopants which end up on Fe-1b sites decrease the energetic cost of the antiferromagnetic stacking of Fe-3e layers and thus effectively increase $J_3/J_2$. On the other hand, the application of pressure mainly affects the interatomic distances, and therefore affects both the ferromagnetic and the antiferromagnetic bonds. The latter is more sensitive to the changing overlap integrals due to the itinerant character of the Ce electrons, thus shifting $J_3/J_2$ towards higher values under pressure. The horizontal scaling of various magnetic phase diagrams for chemical doping and applied pressure results from the fact that there is only a single parameter $J_3/J_2$ which controls the transition between the ferromagnetic and antiferromagnetic phases. By contrast, the thermal transition into the paramagnetic phase is controlled by $J_1$. This strong Ce-Fe spin interaction is little affected by either pressure or doping, and the vertical axes of the different phase diagrams thus line up without any additional scaling.

The picture which emerges for CeFe$_2$ contains multiple magnetic energy scales which compete in a landscape of geometrical frustration, and the magnetic structure is determined by the balance between the weakest two interactions. This solution holds for both chemical substitution and applied pressure and may serve more generally for other geometrically frustrated pyrochlore magnets. The phase transition occurs when the effective interaction between kagome layers turns from ferromagnetic and mediated by the Fe spins, to antiferromagnetic and mediated by the Ce spins. In the ferromagnetic phase, the Fe-Fe interactions are satisfied and the Ce-Ce interactions are frustrated. The rhombohedral distortion tilts the playing field in favor of the Ce-Ce interactions by isolating the 1b-Fe spins, with the result that in the antiferromagnetic phase the Fe-Fe interactions are partially frustrated while the Ce-Ce interactions are mostly satisfied. Underpinning the whole phase diagram is the dominant antiferromagnetic Ce-Fe interaction, which depends on strong hybridization between the itinerant Ce $5d$ and the more localized Fe $3d$ orbitals.

The work at the University of Chicago was supported by NSF Grant No. DMR-0907025. The work at beamline 4-ID-D of the Advanced Photon Source and the Material Science Division of Argonne National Laboratory was supported by the U.S. Department of Energy (DOE) Basic Energy Sciences under Contract No. NE-AC02-06CH11357. PCC's work was supported by the U.S. DOE, Office of Basic Energy Science, Division of Materials Sciences and Engineering. Ames Laboratory is operated for the U.S. DOE by Iowa State University under Contract No. DE-AC02-07CH11358.

\section{Methods}
Single crystals of CeFe$_2$ were grown from a Ce-rich binary melt with initial composition Ce$_{0.6}$Fe$_{0.4}$.  High purity Ce (Ames Lab) and Fe were sealed into a three-cap Ta crucible \cite{28} and subsequently sealed into a silica ampule.  The ampule was heated from room temperature to 1100 $^{\circ}$C over 6 hours, cooled to 950 $^{\circ}$C over 3 hours and then slowly cooled to 700 $^{\circ}$C over 120 hours.  Once at 700 $^{\circ}$C the excess liquid was decanted from the single crystals \cite{28}. X-ray diffraction measurements were performed at beamline 4-ID-D of the Advanced Photon Source. A double-bounce Si (111) monochromator and a pair of Pd mirrors produced a focused beam of 20 keV x-rays free from contamination by higher harmonics. CeFe$_2$ crystals of typical dimensions $70 \times 70 \times 40 {\mu}m^3$ were loaded in the diamond anvil cell (DAC) in an argon atmosphere to avoid oxidation. The pressure medium was a 4:1 (volume) methanol:ethanol mixture. A polycrystalline silver grain was used as an in-situ manometer \cite{21}. A helium-membrane-tuned DAC was used to continuously vary pressure at the cryostat base temperature of 3.5 K. X-ray scattering was carried out in the transmission geometry and within the vertical scattering plane for high momentum space resolution. The use of $70^{\circ}$ full cone Boehler-Almax diamond anvils allowed access to a wide range of reciprocal space. A total of four samples were studied under pressure at 3.5 K. The FWHM of the sample rocking curve never exceeded $0.10^{\circ}$ under pressure.


\begin{thebibliography}{50}
\bibitem {20} Contamination by second harmonic x-rays was eliminated by two Pd mirrors used to focus 20 keV x-rays. Harmonics are monitored with a secondary detection channel set to count 40 keV x-rays on the NaI scintillation detector which showed null. The extremely low intensity of the observed (1/2, 1/2, 1/2)-type peaks (Fig.~\ref{fig3}) rule out the charge scattering contribution to the measured magnetic diffraction peaks.
\end{thebibliography}
\end{document}